\documentclass[aps,prl,twocolumn,superscriptaddress,amsmath,amssymb]{revtex4}
\usepackage{graphicx}
\usepackage{amsmath}

\begin{document}

\title{Quasi-One-Dimensional Dipolar Quantum Gases}

\author{Liming Guan}
\affiliation{Institute for Advanced Study, Tsinghua University,
Beijing, 100084,  P. R. China}

\author{Xiaoling Cui}
\affiliation{Institute for Advanced Study, Tsinghua University,
Beijing, 100084,  P. R. China}

\author{Ran Qi}

\affiliation{School of Physics, Georgia Institute of Technology, Atlanta, Georgia 30332, USA}

\author{Hui Zhai}
\affiliation{Institute for Advanced Study, Tsinghua University, Beijing, 100084,  P. R. China}

\date{\today}
\begin{abstract}
In this letter we consider dipolar quantum gases in a quasi-one-dimensional tube with dipole moment perpendicular to the tube direction. We deduce the effective one-dimensional interaction potential and show that this potential is not purely repulsive, but rather has an attractive part due to high-order scattering processes through transverse excited states. The attractive part can induce bound state and cause scattering resonances. This represents the dipole induced resonance in low-dimension. We work out an unconventional behavior of low-energy phase shift for this effective potential and show how it evolves across a resonance. Based on the phase shift, the interaction energy of spinless bosons is obtained using asymptotic Bethe ansatz. Despite of long-range nature of dipolar interaction, we find that a behavior similar as short-range Lieb-Linger gas emerges at the resonance regime.
\end{abstract}
\maketitle

A major effort in cold atom physics nowadays is to achieve degenerate gases of stable polar molecules, using STIRAP technique \cite{Ni}. Polar molecules possess permanent electric dipole moment, which can be polarized and tuned by external electric field. Their dipole interaction strength can be tuned as strong as comparable to or even larger than the Fermi energy of free gases or the confinement energy in a confined geometry \cite{Houston}. Unlike Coulomb or Van der Waals interaction, dipole interaction is highly anisotropic. Thus, realizing polar molecular gases provides unique many-body systems of strong and anisotropic interactions.

Since ultracold quantum gases are dilute systems, many previous studies of such systems tell us an important lesson that understanding two-body problem is crucial for revealing properties of many-body physics. For instance, for short-range isotropic interaction between atoms, the $1/r$ behavior of short-range two-body wave function leads to universal Tan relations for many-body systems \cite{Tan}; understanding of confinement induced resonance \cite{Olshanii} is the basics for discovering super-Tonks gas \cite{superTonk}. For dipole interaction, in three-dimensional(3D) free space, the two-body problem has been solved by a number of works \cite{LiYou, Blume, Roudnev,Shi}, which reveals dipole induced $s$-wave resonance (DIR). That is to say, although to the first-order of Bohn approximation, dipole interaction has no net effect in the $s$-wave scattering channel, the anisotropic nature of dipole interaction allows coupling to higher partial wave channels, through which a strong effective attractive potential is generated in the $s$-wave channel. Such an attractive potential can support bound state and cause $s$-wave scattering resonance. Benefited from the insight of the two-body solution, intriguing properties of fermion superfluids across a DIR have been studied \cite{Qi,Hu}.

The two-body problem with dipole interaction in confinement geometry has only been studied by few papers \cite{Santos,Recati}. In contrast, there are already quite a few studies of many-body dipolar gases in either 1D or 2D geometry. Many works consider the situation that the dipole moment is perpendicular to the 1D tube. In this situation, the interaction is usually taken as purely repulsive $1/z^3$ potential as expected from the first-order of Bohn approximation \cite{1d}. However, with the experience in 3D case, one may wonder whether this is always a good approximation, in particular, when the dipole interaction is strong enough to compare with the confinement energy.

To address this issue, in this work we first numerically solve two-body problem in 3D with transverse confinement potential, from which an effective 1D scattering potential is deduced. We find that this potential has an attractive part when the inter-particle distance is shorter than the confinement length, aside from a $1/z^3$ repulsive part at long range. We show that such a potential can lead to bound state and resonance. This reveals DIR in confined geometry. Moreover, we find that such a potential gives rise to a low-energy behavior of phase shift as $\cot\delta_k \propto -kD[2\eta+\ln(CkD)]$, where $\eta$ is tunable by the ratio between dipole length $D$ (defined later) and confinement length $a_{\perp}$. All above important information of two-body physics has been overlooked in previous studies. Using the information from two-body problem, we can obtain intriguing interaction effect in a  many-body system of spinless bosons with the help of asymptotic Bethe ansatz. We also find that, nearby resonance, $\eta$ becomes very large and dominates over logarithmic term, the many-body system therefore exhibits a behavior similar as short-range Lieb-Linger gas, despite of the long-rang nature of dipole interaction.

\begin{figure}[t]
\includegraphics[height=2.6 in, width=3.4 in]{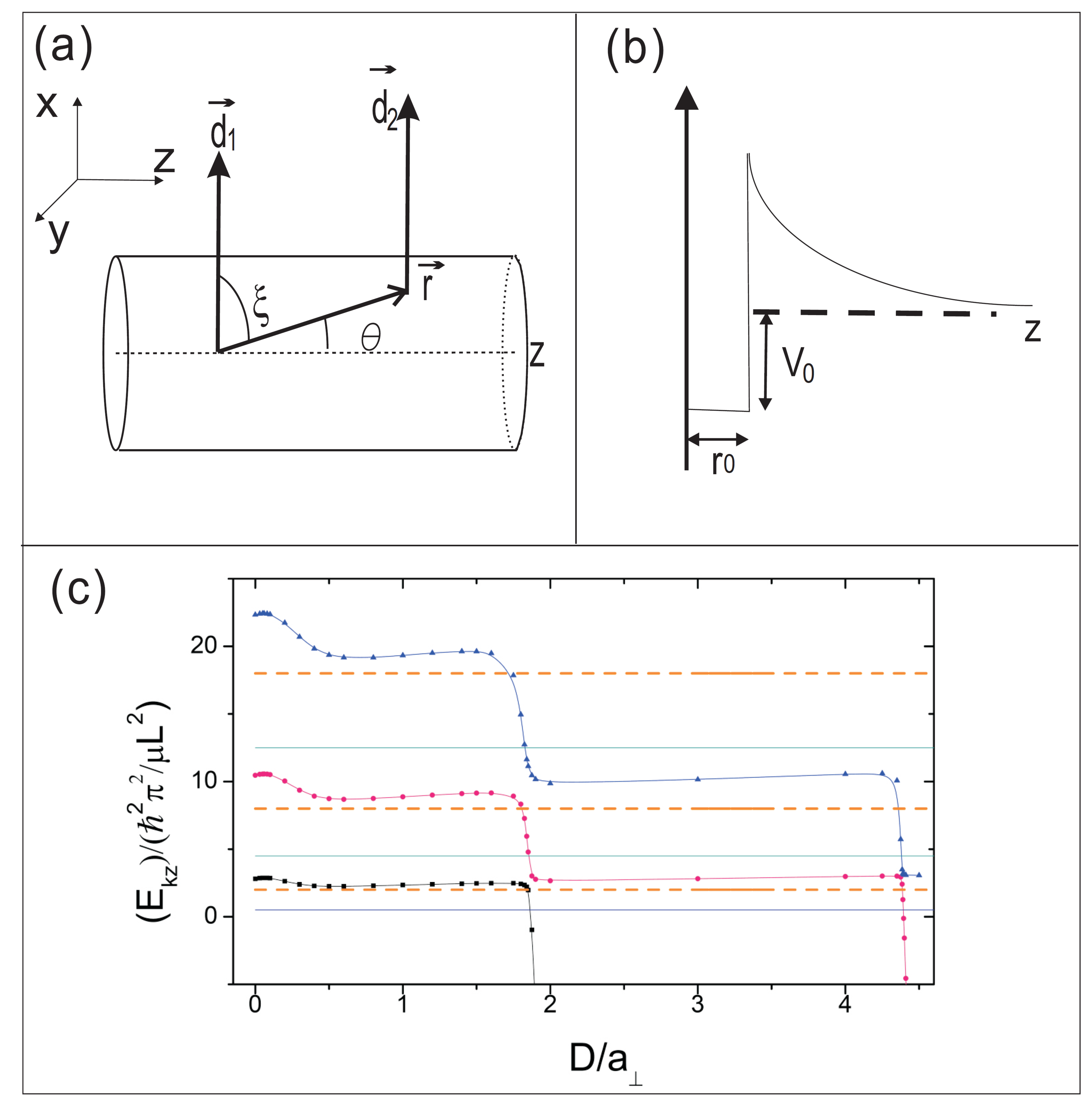}
\caption{(a) Schematic of our system. (b) Schematic of two-body interaction potential. (c) Numerical solution of lowest eigenstates as a function of $D/a_{\perp}$. We take $r_0=0.1a_{\perp}$ and $V_0=-136\hbar\omega$ for the short-range square-well potential. The sample size is $L= 60a_{\perp}$ along z. The solid horizontal line is non-interacting energy level and the dashed horizontal lines are energy level with phase shift $\pi/2$.\label{model}}
\end{figure}

{\it Model.} We consider a one-dimensional system along $\hat{z}$ direction with strong harmonic confinement in the transverse $xy$ plane, and the dipole moment ${\bf d}$ lies in $xy$ plane perpendicular to $\hat{z}$ direction, as shown in Fig \ref{model}(a). The Hamiltonian for the relative motion of two-body problem is given by
\begin{equation}
\Big(-\frac{\hbar^2}{2\mu}\nabla^2+\frac{1}{2}\mu\omega^2\rho^2+V({\bf r})\Big)\psi({\bf r})=E\psi({\bf r}),
\end{equation}
where $\mu=m/2$ is the relative mass, $\rho=\sqrt{x^2+y^2}$ is the transversal radius, $\omega$ is the frequency of the transverse harmonic potential, $V({\bf r})=V_{\text{a}}({\bf r})+V_{\text{d}}({\bf r})$ is shown in Fig. \ref{model}(b). $V_\text{a}$ denotes the potential in atomic scale, and here $V_{\text{a}}({\bf r})$ is chosen as a square well potential $V_{\text{a}}({\bf r})=-V_0$ for $r<r_0$ for simplicity.  $r_0$ is much smaller than confinement length $a_{\perp}=\sqrt{\hbar/\mu\omega}$. The dipole part $V_{\text{d}}({\bf r})=d^2(1-3\cos^2\xi)/r^3$ for $r>r_0$, where $\xi$ is the angle of between ${\bf d}$ and ${\bf r}$. The dipole length $D=\mu d^2/\hbar^2$ can be tuned to be comparable to or even larger than $a_\perp$.

If we fix ${\bf d}$ in the $xy$ plane, say ${\bf d}=d\hat{x}$ (denoted as fixed dipole model), the rotational symmetry along $\hat{z}$ is also broken, which makes the numerical calculation quite involved. To reduce the numerical complicity, we first consider a simpler model, in which ${\bf d}$ is fast rotated along $\hat{z}$ with a frequency much larger than any other energy scales in the problem (denoted by rotating dipole model) as considered in Ref. \cite{Santos}. Upon time average the rotational symmetry along $\hat{z}$ is restored and the effective dipolar interaction becomes $V_{\text{d}}= d^2(3\cos^2\theta-1)/(2r^3)$, where $\theta$ is the angle between ${\bf r}$ and $\hat{z}$. With the knowledge obtained from the rotating dipole model, our discussion will later return to the fixed dipole model since it is more realistic.

{\it Numerical Results of Rotating Dipole Model.} We numerically solve the rotating dipole model with discrete variable representation \cite{Baye, Szalay,Tiesinga} in a cylindrical box. We consider $s$-wave scattering and the corresponding even-parity eigen states. Thus, in the rest of this paper we only wave function at $z>0$. The eigen-spectrum is plotted as a function of $D/a_{\perp}$ in Fig \ref{model}(c). We find that as $D/a_{\perp}$ increases, a series of bound states appear in sequence.  We have also found that the positions for the onset of bound states are insensitive to changing $V_0$ with $r_0$ fixed, which only modifies  the atomic potential part. Rather, the onsets of bound state are sensitive to changing $r_0$ with $V_0$ fixed, which modifies not only  the atomic potential but also the cutoff length scale of dipole part. This strongly  indicates that the bound state originates from dipolar interaction rather than the potential at atomic scale.

%\begin{figure}[t]
%\includegraphics[height=2.5 in, width=3.2 in]
%{Fig2}
%\caption{Level spectrum. (a) $r_0/a_\perp=0.1$ and $V_0=?$; (b) $r_0/a_\perp=0.1$ and $V_0=?$ (c) $r_0/a_\perp=0.15$ and $V_0=?$ (d) $r_0/a_\perp=0.15$ and $V_0=?$ \label{spectrum}}
%\end{figure}

Therefore, to understand how dipole interaction could induce bound states, we shall deduce an effective one-dimensional interaction potential $V_{\text{1d}}(z)$. Previously, one commonly approach is to assume that molecules always stay in the lowest transverse confinement mode $\phi_0(\rho)$, and this single mode approximation (SMA) gives
\begin{align}
&V_{\text{1d}}^{\text{SMA}}\left(\frac{z}{a_\perp}\right)=\int\int dxdy V_{\text{d}}({\bf r})\phi^2_0(\rho)\nonumber \\
&=\frac{\hbar^2}{\mu} \frac{D}{a_{\perp}^3}\left[\sqrt{\pi}\left(1+2\frac{z^2}{a^2_\perp}\right)e^{\frac{z^2}{a^2_\perp}}\text{erfc}\left(\frac{z}{a_\perp}\right)-2\frac{z}{a_\perp}\right].
\end{align}
As expected, $V_{\text{1d}}^{\text{SMA}}(z)$ is a purely repulsive potential and behaves as $(\hbar^2/\mu)D/z^3$ for large $z/a_\perp\gtrsim1$, and this potential can not support any bound state. Alternatively, we can deduce $V_{\text{1d}}(z)$ from the numerical solution of eigenfunction $\psi(\rho,z)$ and eigenvalue $E$  in the following way. Noting that the overall weight of $\psi(\rho,z)$ on the lowest transverse mode $\phi_0(\rho)$ is nearly unity in a quite wide range of $z/a_{\perp}$ (see the inset of Fig.2(a)), we can therefore define a projected 1D wave function
%since now we have already obtained the eigenfunction $\psi(\rho,z)$ and eigenvalue $E$ numerically, we can deduce $V_{\text{1d}}(z)$ as follows: by projecting $\psi(\rho,z)$ to $\phi_0(\rho)$ mode we have
$\psi_{\text{1d}}(z)=\int dxdy \phi_0(\rho)\psi(\rho,z)$. Assuming $\psi_{\text{1d}}(z)$  satisfies a 1D Schr\"odinger equation $[-(\hbar^2/2\mu) d^2/dz^2+V_{\text{1d}}(z)]\psi_{\text{1d}}(z)=(E-\hbar\omega)\psi_{\text{1d}}(z)$, we obtain $V_{\text{1d}}$ as
\begin{equation}
V_{\text{1d}}(z)=\frac{\hbar^2}{2\mu} \frac{\frac{d^2 }{dz^2} \psi_{1D}(z)}{\psi_{1D}(z)}+E-\hbar\omega,
\end{equation}

\begin{figure}[bp]
\includegraphics[height=1.7 in, width=3.4 in]{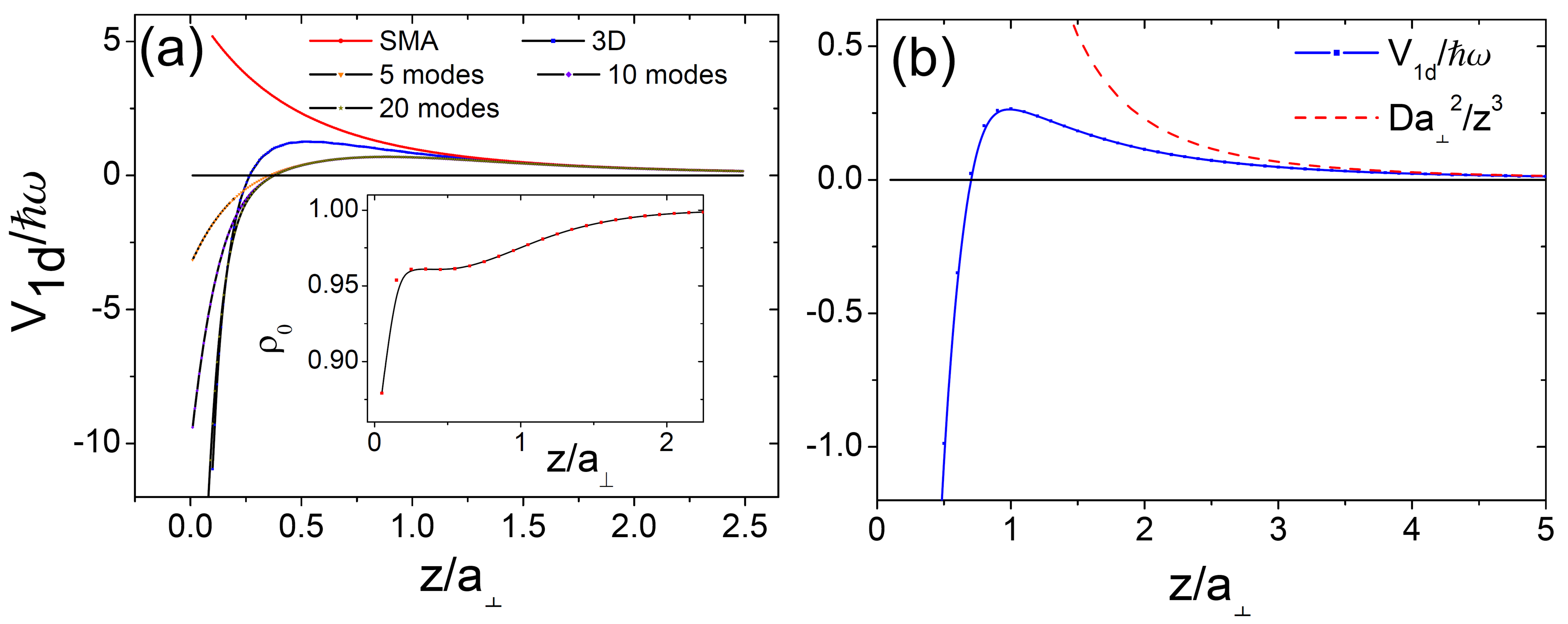}
\caption{Effective potential for rotating dipole model (a) and fixed dipole model (b) at $D/a_{\perp}= 1.825$. Inset of (a) shows the weight of $\psi(\rho,z)$ on the lowest transverse mode, i.e., $\rho_0=\frac{|\psi_{1d}(z)|^2}{\int |\psi(\rho,z)|^2 d^2\rho}$. Different lines in (a) correspond to SMA, MMA with $5$, $10$ and $20$ modes included, and full three-dimension numerical results. In (b), blue sold line corresponds to MMA with $10$ modes, and the dashed line in (b) corresponds to $D/z^3$ potential.  \label{potential}}
\end{figure}

In Fig. \ref{potential}(a) we compare $V_{\text{1d}}$ with $V^{\text{SMA}}_{\text{1d}}$ and find that in the regime $z/a_{\perp}\gtrsim1$ these two potentials agree very well. While at short distance when $z/a_{\perp}\lesssim1$, $V_{\text{1d}}$ starts to deviates from $V^{\text{SMA}}_{\text{1d}}$ and finally $V_{\text{1d}}$ becomes attractive. This is because dipole interaction increases strongly at short distance and can overcome the confinement potential, and the higher-order scattering processes via higher transverse modes give rise to attraction. This mechanism is in fact identical to DIR in 3D. Thus, when the short range attraction in $V_{\text{1d}}$ leads to a bound state nearby threshold, the low-energy scattering behavior will be dramatically modified and $V^{\text{SMA}}_{\text{1d}}$ fails to describe the low-energy physics.

Nevertheless, the SMA can be improved by including multi-modes. Considering $\hat{H}_{\rho}=-\frac{\hbar^2}{2\mu}(\partial^2_x+\partial^2_y)+\frac{1}{2}\mu\omega^2\rho^2+V({\bf r})$ and expanding $\hat{H}_{\rho}$ in the harmonic oscillator eigen-basis $\phi_{i}(\rho)$ as $H^{ij}_{\rho}(z)=\int dxdy \phi_{i}(\rho)\hat{H}_{\rho}\phi_{j}(\rho)$, the matrix $H_\rho(z)$ can be diagonalized by a unitary transformation $X^\dag(z)H_{\rho}(z)X(z)=\Lambda(z)$. By adiabatic approximation \cite{Shi}, the multi-mode approach (MMA) gives an effective potential  $V^{\text{MMA}}_{\text{1d}}(z)=\Lambda^{00}(z)$. In Fig 2(a) we also compare $V^{\text{MMA}}_{\text{1d}}(z)$ with $V_{\text{1d}}$ and we find that $V^{\text{MMA}}_{\text{1d}}(z)$ can reproduce the short-range attractive behavior reasonably well, as long as one keeps sufficient modes. Thus, our numerical results show that the MMA captures main features in 1D effective potential. Therefore, we have confidence that the MMA can also be applied to the fixed dipole model. An effective potential for fixed dipole model is obtained as shown in Fig. \ref{potential}(b) with MMA. Similar as the rotating dipole model, it coincides with $D/z^3$ when $z\gtrsim a_\perp$ and gradually becomes attractive when $z\lesssim a_\perp$. Our following discussion is based on such an effective 1D potential, which can be applied to both two models.

{\it Low-Energy Phase Shift.}  When $z\gtrsim a_\perp$, the wave functions can be solved perturbatively in two different regimes. In regime I, $z\gg D$, and in regime II, $z\ll 1/k$. The wave function for $z\lesssim a_\perp$, where the potential deviates from $D/z^3$,  will not be studied explicitly; instead, it determines the boundary condition for regime II.

In regime I, we take $\psi_I(z)=A_k(\cot\delta_k W(kz)- V(kz))$, where $A_k$ is a normalization factor. $W(kz)$ and $V(kz)$ are two independent solutions and both can be expanded as $W(kz)=\sum_n(kD)^nW_n(kz)$ and $V(kz)=\sum_n(kD)^nV_n(kz)$ when $D/z^3$ term is treated as a perturbation for $z\gg D$.  To the zeroth order we have $\psi_I(z)\propto\cos(kz+\delta_k)$, and thus,
$W_0(kz)=\cos(kz)$ and $V_0(kz)=\sin(kz)$ where $\delta_k$ is the phase shift. To the next order, we find $W_1(\xi)$ and $V_1(\xi)$ ($\xi$ denotes $kz$) as
\begin{align}
W_1(\xi)&=-\text{Ci}(2\xi)\sin\xi+\frac{1}{2}\cos\xi\Big[\frac{1}{\xi}+2\text{Si}(2\xi)-\pi \Big]\nonumber\\
V_1(\xi)&=-\text{Ci}(2\xi)\cos\xi+\frac{1}{2}\sin\xi\Big[\frac{1}{\xi}-2\text{Si}(2\xi)+\pi \Big]
\end{align}
where $\text{Ci}(z)=-\int_z^{\infty}\frac{\cos t}{t} dt$ and $\text{Si}(z)=\int_0^z\frac{\sin t}{t} dt$. Up to the order of $O(\xi)$, by expanding $\psi_I(z)$ in terms of $\xi$, we have
%\begin{align}
%&\psi_{I}(z)=A_k\left\{-kD\tan\delta_k \frac{z}{D}-kD\tan\delta_k\ln(z/D)+1\nonumber\right.\\
%&\left.-\frac{kD}{2}(\pi+\tan\delta_k(1-2\gamma_E))+kD\tan\delta_k\ln(2kD)\right\} \label{psi_I}
%\end{align}
\begin{align}
&\psi_{I}(z)=A_k\left\{\cot\delta_k-kD\frac{z}{D}\nonumber\right.\\
&\left.-\frac{kD}{2}\cot\delta_k[1-2\gamma_E-2\big(\ln(2kD)+\ln(2z/D)\big)]\right\},  \label{psi_I}
\end{align}
where $\gamma_\text{E}$ is Euler's constant.

\begin{figure}[tb]
\includegraphics[height=2.0 in, width=3.2 in]{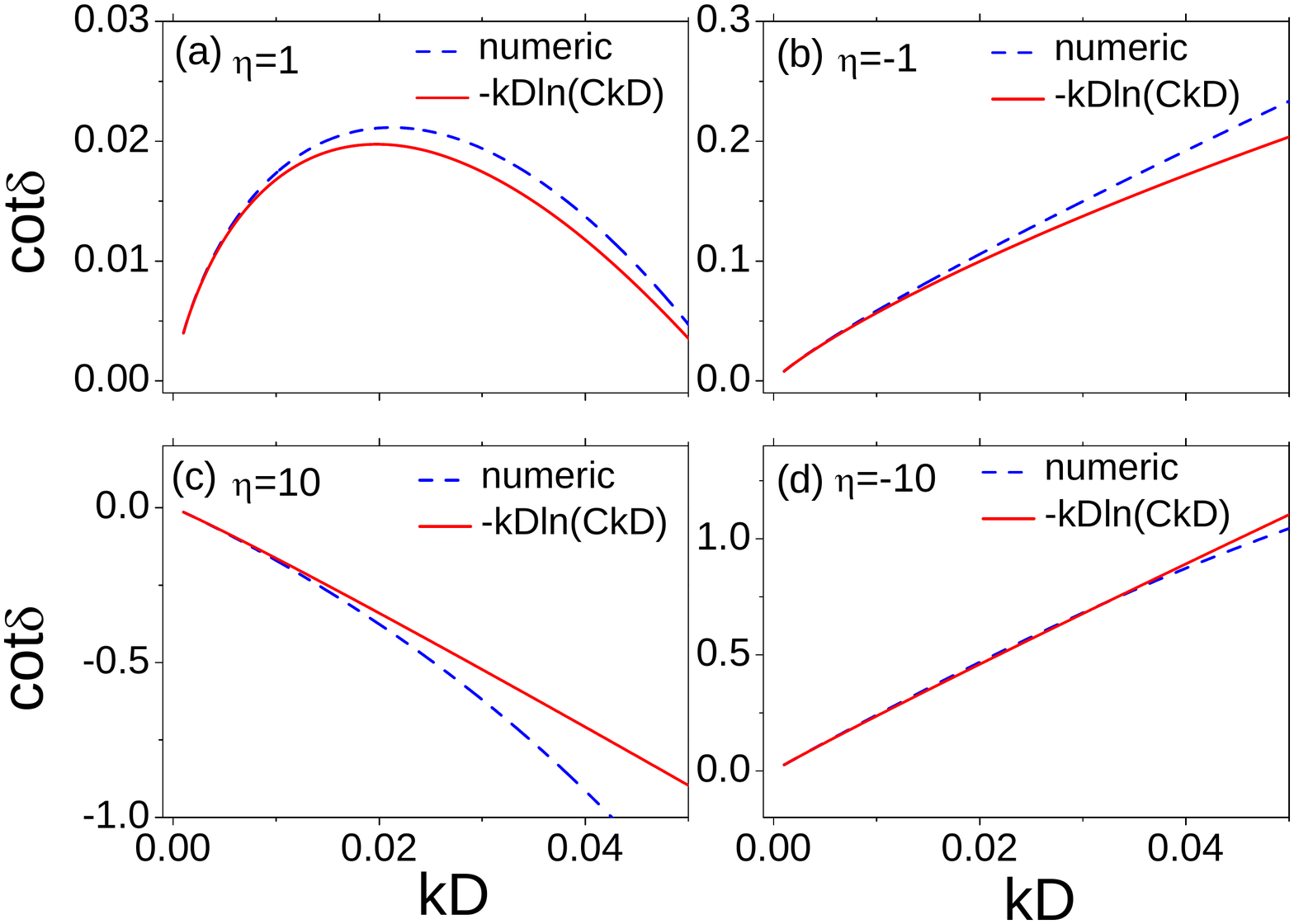}
\caption{$\cot\delta_k$ determined by numerics (blue dashed lines) are compared with analytical formula Eq.8 (red solid lines) for $\eta=1$(a), $\eta=-1$(b), $\eta=10$(c) and $\eta=-10$(d). \label{phase_shift}}
\end{figure}

In regime II, when $z\ll 1/k$, we can treat $k^2\psi(z)$ term as a perturbation. To the order of $O(kz)$, we only need to consider the zero-energy solutions of $D/z^3$ potential, with general form given by
\begin{eqnarray}
\psi_{II}(z)=\sqrt{z/D}[K_1(2\sqrt{D/z})+\eta I_1(2\sqrt{D/z})],
\end{eqnarray}
where $K_1$ and $I_1$ are respectively the regular and irregular solutions for $D/z^3$ potential. Their relative coefficient $\eta$ is determined by potential details at short range, which can be tuned by $D/a_\perp$ in the present model (see the inset of Fig.4). In following discussion of low-energy physics, $\eta$ serves as an independent input parameter, which describes the effect of short-range potential to the long-range physics.
%Here we shall treat $\eta$ as an independent parameter tunable by $D/a_\perp$. In another word, $\eta$ appears in following discussion of low-energy physics as a parameter independent of $D$, which replaces the role of $D/a_\perp$ in the original three-dimensional model.
If the potential is close to a pure $D/z^3$ potential, $\eta\rightarrow 0$; if the short-range part is about to form a bound state, $\eta\rightarrow+\infty$; and if a bound state has been formed nearby threshold, $\eta\rightarrow -\infty$.
%, as shown in the inset of Fig. \ref{1dboson}.
By expanding $\psi_{II}(z)$ at large $z/D$, one obtains
\begin{equation}
\psi_{II}(z)=\frac{z}{2D}-\frac{\ln (z/D)}{2}+\eta+\gamma_\text{E}-\frac{1}{2}+O(D/z). \label{psi_II}
\end{equation}

Finally for low-energy scattering $kD\ll 1$, we require $\psi_I(z)$ of Eq. \ref{psi_I} and $\psi_{II}(z)$ of Eq. \ref{psi_II} behave in the same way in their overlap region $D\ll z\ll 1/k$. Thus, by equating the relative coefficients of $z/D$, $\ln(z/D)$ and constant terms between Eq. \ref{psi_I} and Eq. \ref{psi_II}, we can obtain two coupled equations for $A_k$ and $\cot\delta_k$, the solution of which gives the low-energy behavior of phase shift
\begin{eqnarray}
\cot\delta_k=-kD[2\eta+\ln(CkD)], \   \  C=2e^{3\gamma_E-3/2}. \label{pshift}
\end{eqnarray}

To verify this phase shift we numerically solve the 1D Schr\"odingier equation with a boundary condition at $z^*\ll D$, and the boundary condition can be satisfied by a given $\eta$ for zero-energy solution. From the numerical solution we can extract $\delta_k$ from wave function behavior at large distance. We plot $\cot\delta_k$ as a function of $kD$ and compare it with the formula Eq. \ref{pshift} in Fig. \ref{phase_shift}, and find that they agree reasonably well for sufficient low-energies \cite{better}. The logarithmic correction reflects the long-range nature of dipole interaction.

\begin{figure}[tb]
\includegraphics[height=2.2 in, width=3.2 in]{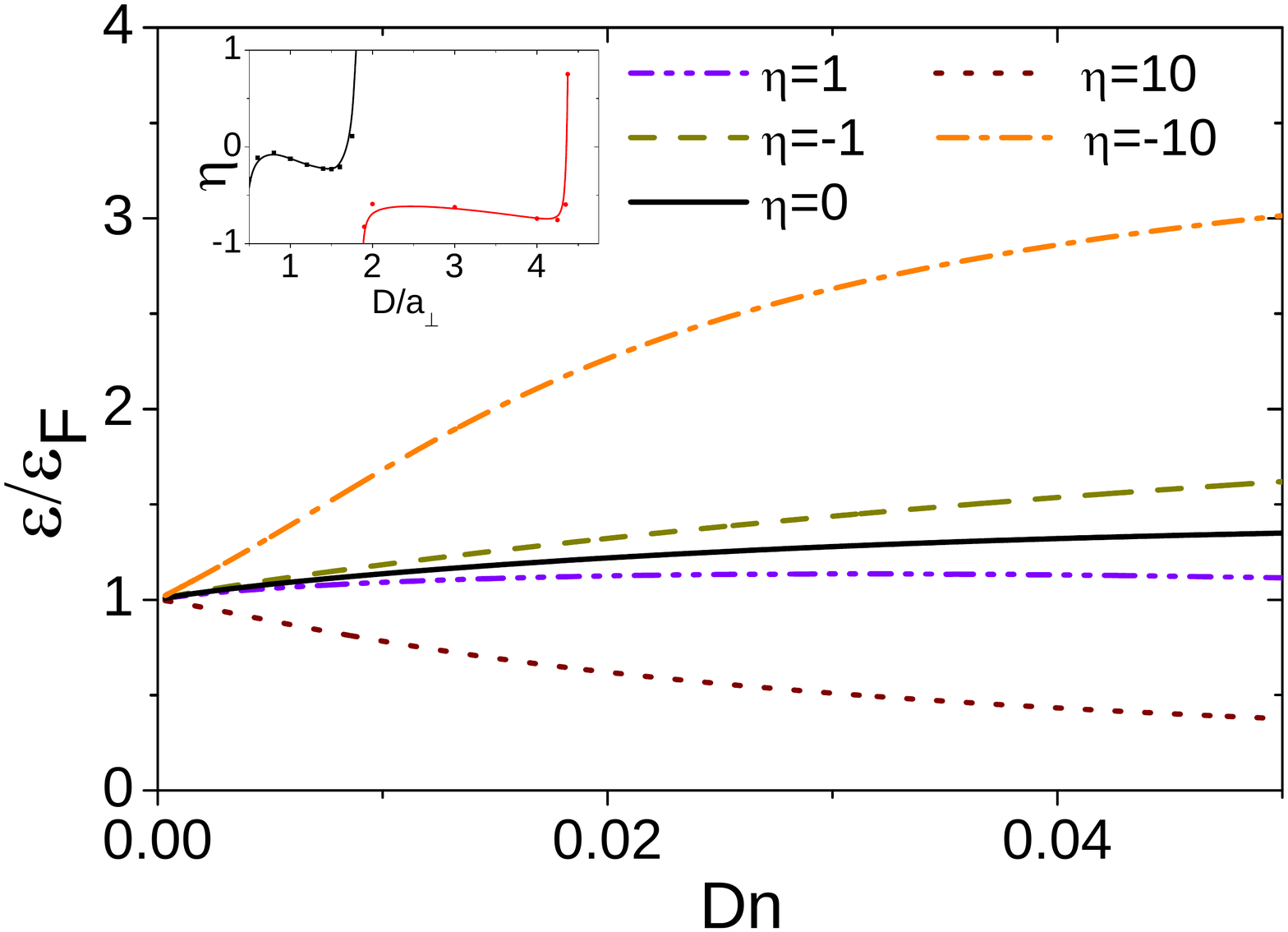}
\caption{Energy density of spinless bosons as a function of $Dn$ for different $\eta$. $n$ is the density of 1D bosons. $\mathcal{E}_F$ is the energy density for identical fermions with the same density. $\eta$ can be tuned by $D/a_{\perp}$ in the quasi-1D model, as shown in the inset.  \label{1dboson}}
\end{figure}

{\it Many-body System of Spinless Bosons.} Hereafter we study a many-body system of spinless dipolar bosons with asymptotic Bethe Ansatz \cite{ABA}.
Asymptotic Bethe Ansatz only makes use of the scattering phase shift to obtain the thermodynamics of a system with finite density. Following the standard procedure \cite{ABA}, we obtain $k_j L=2\pi I_j -2\sum_{i\neq j} \delta((k_j-k_i)/2)$, where $I_j$ are quantum numbers and  $\delta_k$ is the two-body phase shift obtained in Eq. \ref{pshift}. In the thermodynamic limit one obtains a Fredholm type equation
\begin{equation}
\rho(k)=\frac{1}{2\pi}+\frac{1}{\pi}\int_{-B}^{B}\frac{\partial \delta(\frac{k-q}{2})}{\partial k}\rho(q)dq. \label{ABA}
\end{equation}
The density and the energy density are
\begin{equation}
n=\int_{-B}^{B}\rho(k)dk,  \quad\quad\quad \mathcal{E}=\int_{-B}^{B}k^2\rho(k)dk. \label{density},
\end{equation}
The solution of Eq. \ref{ABA} and Eq. \ref{density} gives $\mathcal{E}/\mathcal{E}_\text{F}$ as a function of $Dn$ for different $\eta$. In Fig. \ref{1dboson} we display how energy density behaves across a resonance.

We find for all $\eta$, the energy density approaches the Tonks limit, i.e. $\mathcal{E}/\mathcal{E}_\text{F}\rightarrow 1$, in the dilute limit of $Dn\rightarrow 0$, because phase shift $\delta\rightarrow \pi/2$ when $k\rightarrow 0$. Nearby a resonance when $|\eta|$ is sufficiently large, as long as $kD\gg e^{-|\eta|}$, $2\eta$ term dominates over the logarithmic term in Eq. \ref{pshift}, and the phase shift can be well approximated as $\cot\delta_k=-kD\eta$. This gives a nearly energy-independent interaction constant $g_\text{1d}=\hbar^2/(\mu D\eta)$, and the interaction parameter $\gamma=1/(\eta nD)$.
% , and  For positive $\eta$, similar as usual Lieb-Linger gas \cite{LL}, the ratio between interaction energy and kinetic energy $\gamma=1/(\eta nD)$. Thus, as shown in Fig. \ref{1dboson}, energy density decreases as $nD$ increases. For negative $\eta$, this system is similar as in the super-Tonk regime, and energy density increases with $nD$ \cite{superTonk_theory}.
$\gamma$ decreases with $D$ for $\eta>0$ while increases for $\eta<0$. As shown in Fig.4, for positive $\eta$, the energy density $\mathcal{E}$ decreases as $D$ increases, similar to usual Lieb-Liniger gas \cite{LL}; for negative $\eta$, this system is in the super-Tonks regime, and $\mathcal{E}$ increases with $D$ \cite{superTonk_theory}.
Finally, once away from resonance, when $|\eta|$ is small, the logarithmic term dominates over $2\eta$ term in Eq. \ref{pshift}, as long as $kD\ll 1$. In this regime, the physics can not be analogous to a gas with short-range contact interaction, since the logarithmic term prevents defining an energy independent interaction constant. In this regime, numerically we find  energy density varies much slowly with $Dn$.

{\it Final Comment:} Our discussion here can be straightforwardly generalized to the quasi-2D case, where a confinement potential is applied along $\hat{z}$ direction and dipole moment is perpendicular to $xy$ plane (along $\hat{z}$). In this case, the 2D effective potential also contains a long range $1/\rho^3$ part and a short range attractive part. However, compared to quasi-1D case, in quasi-2D it is more difficult for such a potential to form a bound state. In another word, one needs a much larger $D/a_z$, where $a_z$ is the confinement length along $\hat{z}$, to facilitate a stronger attraction. With larger $D/a_z$, the repulsive barrier is also much higher and the coupling between bound state and scattering state is much weaker \cite{zheyu}. Hence, the resonance is extremely narrow in quasi-2D. That is to say, the irregular solution is not important ($\eta\approx 0$) except for very narrow range of $D/a_z$. A pure $1/\rho^3$ repulsive potential works much better in 2D comparing to 1D case discussed in this paper.

{\it Acknowledgements.} We thank Gora Shlyapnikov, Peng Zhang and Zhe-Yu Shi for useful discussion. This work is supported by Tsinghua University Initiative Scientific Research Program, NSFC under Grant No. 11104158 (XC),  No. 11104157(RQ), No. 11004118 (HZ), No. 11174176 (HZ), and NKBRSFC under Grant No. 2011CB921500.


\begin{thebibliography}{99}
\bibitem{better} We find an empirical fitting formula of phase shift $\cot\delta_k=-kD[2+(1/\eta)(\ln(CkD)]/(1/\eta+b (kD)^2\ln(kD))
$ which fits the numerically results in wider energy range, in particular, in the regime when $\eta$ is large. Here b is a fitting constant. For small $\eta$ or small $k$, this formula is consistent with Eq. \ref{pshift}.
\end{thebibliography}
\end{document}